\documentclass[nofootinbib]{revtex4}
\usepackage{graphicx,amsfonts}

\begin{document}

%



\title{The group field theory approach to quantum gravity}

\author{{\bf Daniele Oriti}}

\affiliation{Department of Applied Mathematics and Theoretical Physics \\ Centre
for Mathematical Sciences, University of Cambridge \\ Wilberforce
Road, Cambridge CB3 0WA, England \\ d.oriti@damtp.cam.ac.uk}

\begin{abstract}We give a very concise review of the group field
  theory formalism for non-perturbative quantum gravity, a higher
  dimensional generalisation of matrix models. We motivate it as a
  simplicial and local realisation of the idea of 3rd quantization
  of the gravitational field, equivalently of a quantum field theory of
  simplicial geometry, in which also the topology of space is fully dynamical. We highlight the basic
  structure of the formalism, and discuss briefly various models that
  are being studied, some recent results and
  the many open issues that future research should
  face. Finally, we point out the connections with other approaches to quantum gravity, such as
  loop quantum gravity, quantum Regge calculus and dynamical
  triangulations, and causal sets.\end{abstract}
\maketitle
\section{Introduction and motivation}
Group field theories (GFTs) \cite{iogft,laurentgft} were developed at first as a
generalisation of matrix models for 2d quantum gravity to 3 and 4
spacetime dimensions to produce a lattice formulation of topological
theories. More recently, they have been developed further in the
context of spin foam models for quantum gravity, as a tool to overcome
the limitations of working with a fixed lattice in the
non-topological case. In our opinion,
however, their role in quantum gravity
research goes much beyond this original scope, and GFTs should be seen
as a fundamental formulation of
quantum gravity and not just as an auxiliary tool.  The bottom line of
this perspective, here only tentatively outlined and
still to be fully realised, hopefully, after much more work, can be summarised
as follows: GFTs are quantum field theories {\bf of} spacetime (as
opposed to QFTs {\it on} spacetime), that describe the dynamics of
both its topology and geometry in local, simplicial, covariant, algebraic
terms, and that encompass most of the other approaches to
non-perturbative quantum gravity, providing a link between them and at
the same time going beyond the limitations of each.
It will be clear from this brief account that we have just began to
explore the structure of these models, and that very little is known
about them; at the same time, we hope it would also be clear that what
is known is enough for suggesting that in the GFT framework lies the
potential for important developments towards a complete
theory of quantum gravity.

\
\

The idea of defining a quantum field theory {\it of} geometry, i.e. a QFT on superspace (the
space of 3-geometries) for given spatial topology, say $S^3$, was already
explored in the past \cite{giddstro,banks,mcguigan}. The context was then a \lq quantum cosmology\rq one, i.e. a global one. Such a theory would produce, in its perturbative expansion, a sum over different
topologies each corresponding to a possible Feynman graph of the
theory and to a possible interaction process for \lq universes\rq
represented by the basic 3-sphere. The spatial topology change would
be limited therefore to a changing number of disjoint copies of $S^3$. The field would represent a 2nd quantisation of the canonical wave function
on superspace, here describing the \lq\lq one-particle sector\rq\rq of
the theory\footnotemark \footnotetext{The 3-metric being itself a field, this 2nd
  quantisation of what is already a field theory was dubbed, with a
  certain abuse of language, a 3rd quantisation.}, turning it into an
operator. The quantum amplitude for each Feynman graph, corresponding
to a particular spacetime topology with $n$ boundary components, would
be given by a path integral for gravity on the given topology, i.e. by
a sum-over-histories quantisation of gravity with amplitude for each
history obtained by exponentiating ($i$ times) the Einstein-Hilbert action for
General Relativity. The difficulties in making mathematical sense of
the continuum path integral itself are well-known, and it is a
safe guess that the technical difficulties in turning this 3rd quantisation
idea in a mathematically rigorous framework in the continuum are even
more formidable. Also, such a cosmological setting (having a wave
function and then a field for the whole universe) presents notorious
interpretation problems, here only enhanced by treating also space
topology as a variable) and lead to more than a few worries about the
physical meaning of the whole setting. The general idea however is
appealing in that such a framework would provide a natural mechanism
for implementing topology change in quantum gravity and a prescription
for weighting different spacetime topologies within a covariant
sum-over-histories quantisation of gravity. In particular one could
imagine that the interpretation issues, if not the technical
difficulties, would be made easier if it was possible to implement the
above ideas in a {\bf local} framework, for example by generalising
the superspace construction to {\it open chunks of the universe}, for
example 3-balls, and then describing in a 3rd quantised language the
interaction of these local pieces of the universe generating
dynamically the whole universe and spacetime through their
evolution. Again, however, the continuum setting seems to prevent a rigorous
realisation of these ideas. By turning to a
{\bf simplicial} description of spacetime, the group
field theory formalism gives a mathematically better defined realisation of these appealing ideas, providing a discrete
sum-over-histories quantisation of gravity, and allowing an easier
physical interpretation, since the picture of geometry and
topology it is based on is intrinsically {\it local}.

\section{The general formalism}
\subsection{Kinematics}
The geometry of a simplicial space (a triangulation) is fully characterised by a
countable, if not finite, number of variables only, i.e. superspace becomes
discrete, thus facilitating greatly the task of defining a field theory on it. Also, it is known that every closed
D-dimensional simplicial complex can be obtained by gluing fundamental
D-dimensional building blocks, each with the topology of a D-ball, along
their boundaries (given by (D-1)-simplices). A {\bf local}, and thus
more physically sensible, realisation of the idea of a field theory on
superspace is then possible, by considering first a wave function associated to each (D-1)-dimensional simplicial building block of space (if spacetime is D-dimensional), i.e. a
  functional of the geometry of one of them only, and then second
  quantising it, i.e. turning it into an operator. The quantum
  geometry of larger blocks of a spatial simplicial complex will be encoded in
  the tensor product of such wave functions/operators for the
  individual building blocks forming them. The issue is then: how to
  characterise the geometry of each simplicial building block, and
  thus of the full simplicial complex? in other words, which variables
  to choose to represent the
  arguments of our wave function/field? Here group field theories
  differ from other simplicial approaches to quantum gravity, and follow instead the path traced by loop quantum gravity, in that
  they seek to describe quantum geometry in terms of algebraic data,
  i.e. group and representation variables. This descends \cite{iogft,
    review, alexreview} from the classical description of gravity in terms of
  connection variables valued in the Lie algebra of the Lorentz group
  of the appropriate dimension, discretised to give elementary group
  valued parallel transports along paths in the (dual of the)
  simplicial complex, or equivalently in terms of Lie algebra-valued (D-2)-forms as in
  BF-like formulations of the same \cite{Peldan,FKP}, discretized to
  give the volumes of the (D-2)-dimensional cells of the simplicial
  complex, labelled by irreducible representations of the Lorentz
  group. The equivalence between these two sets of variables is given
  by the harmonic analysis on the group manifold that expresses their
  conjugate nature. More concretely, the field is taken to be a
  $\mathbb{C}$-valued function of D group elements, for a generic
  group $G$, one for each of the D boundary (D-2)-faces of the
  (D-1)-simplex the field corresponds to: $$\phi(g_1,g_2,...,g_D):
  G^{\otimes D}\rightarrow \mathbb{C}.$$ The order of the arguments in
  the field corresponds to a choice of orientation for the (D-1)-simplex
it represents; therefore it is natural to impose the field to be
invariant under even permutations of its arguments (that do not change
the orientation) and to turn into its own complex conjugate under odd
permutations; this choice ensures that only orientable complexes are
generated in the Feynman expansion of the field theory \cite{DP-P}. Other symmetry properties can also be considered
\cite{DP-F-K-R}. The closure of the D (D-2)-faces to form a
(D-1)-simplex is expressed algebraically by the invariance of the field
under diagonal action of the group $G$ on the D arguments of the
field: $\phi(g_1,...,g_D)=\phi(g_1g,...,g_Dg)$, which is also
imposed \cite{BB,carlomike}. This is also the simplicial counterpart
of the Lorentz gauge invariance of first order gravity actions. As anticipated, the field can be expanded in modes using
harmonic analysis on the group:
$$\phi(g_i)=\sum_{J_i,\Lambda,k_i}\phi^{J_i\Lambda}_{k_i}\prod_iD^{J_i}_{k_il_i}(g_i)C^{J_1..J_4\Lambda}_{l_1..l_4}, $$ with the $J$'s
labelling representations of $G$, the $k$'s vector
indices in the representation spaces, and the $C$'s being intertwiners
of the group $G$, an orthonormal basis of which is
labelled by an extra parameter $\Lambda$. That this decomposition is
possible is not guaranteed in general but it is in fact true for the
models being developed up to now, all based on the Lorentz group or on
extensions of it. Group variables represent thus the GFT analogue of
configuration space, while the representation parameters label the
corresponding momentum space. Geometrically, the group variables, as
said, represent parallel transport of a connection along elementary
paths dual to the (D-2)-faces, while the representations $J$ can be
put in correspondence with the volumes of the same (D-2)-faces, the
details of this correspondence depending on the specific model \cite{review,alexreview,thesis}. The
1st quantisation of a geometric (D-1)-simplex in terms of these
variables was performed in great detail in the 3 and 4-dimensional
case in \cite{BB}, but a similar analysis is lacking in higher
dimensions. A simplicial space built out of $N$ such (D-1)-simplices
is then described by the tensor product of $N$ such wave functions, at
the 1st quantised level, with suitable constraints implementing their
gluing, i.e. the fact that some of their (D-2)-faces are identified,
at the algebraic level. For example, a state describing two
(D-1)-simplices glued along one common (D-2)-face would be represented
by: $\phi^{J_1 J_2..J_D\Lambda}_{k_1 k_2...k_D} \phi^{\tilde{J}_1
  J_2...\tilde{J}_D\tilde{\Lambda}}_{\tilde{k}_1 k_2...\tilde{k}_D}$,
where the gluing is along the face labelled by the representation
$J_2$, and effected by the contraction of the corresponding vector
indices (of course, states corresponding to disjoint (D-1)-simplices
are also allowed). The corresponding state in configuration variables
is: $\int dg_2
\phi(g_1,g_2,...,g_D)\phi(\tilde{g}_1,g_2,...,\tilde{g}_D)$. We see that states of the theory are then labelled, in momentum space, by {\it spin networks} of the group $G$ (see chapters by Thiemann and Perez). The
second quantisation of the theory is obtained by promoting these wave
functions to operators, as said, and the field theory is then
specified by a choice of field action and by the definition of the
partition function of the corresponding quantum theory. The partition
function is then expressed perturbatively in terms of Feynman
diagrams, as we are
going to discuss. This implicitly assumes a description of the
dynamics in terms of creation and annihilation of (D-1)-simplices,
whose interaction generates a (discrete) spacetime as a particular
interaction process (Feynman diagram) \cite{DP-F-K-R}. This picture has not
been worked out in detail yet; in other words, no clear Fock structure
on the space of states has been constructed, and no explicit
creation/annihilation operators, acting on a suitably \lq third
quantised vacuum\rq, resulting from the classical
Hamiltonian structure of a group field theory, have been defined. Work
on this is in progress \cite{ahmed-io-jimmy} (see also \cite{mikovic}).

\subsection{Dynamics}
On the basis of the above kinematical structure, spacetime is viewed
as emerging from the interaction of fundamental building blocks of
space, and inherits therefore their simplicial characterisation, in
that it is represented by a D-dimensional simplicial complex. More
precisely, any such simplicial spacetime is understood as a particular
interaction process among (D-1)-simplices, described in term of
Feynman diagrams in typical quantum field-theoretic perturbative
expansion. With this in mind, it is easy to understand the choice of
classical field action in group field theories. This action, in
configuration space, has the general QFT structure:

\begin{eqnarray} S_D(\phi, \lambda)= \frac{1}{2}\left(\prod_{i=1}^D\int
  dg_id\tilde{g}_i\right)
  \phi(g_i)\mathcal{K}(g_i\tilde{g}_i^{-1})\phi(\tilde{g}_i)
  +
  \frac{\lambda}{(D+1)!}\left(\prod_{i\neq j =1}^{D+1}\int dg_{ij}\right)
  \phi(g_{1j})...\phi(g_{D+1 j})\,\mathcal{V}(g_{ij}g_{ji}^{-1}), \label{eq:action}
\end{eqnarray}
where the choice of kinetic and interaction functions $\mathcal{K}$
and $\mathcal{V}$ define the specific model considered. Of course, the
same action can be written in momentum space after harmonic
decomposition on the group manifold. The combinatorial structure of
the dynamics is evident from the pairing of the arguments of these two
functions and from their degree in field variables: the interaction
term describes the interaction of D+1 (D-1)-simplices to form a
D-simplex by gluing along their (D-2)-faces (arguments of the fields),
that are pairwise linked by the interaction vertex. The nature of this
interaction is specified by the choice of function $\mathcal{V}$. The
(quadratic) kinetic term involves two fields each representing a given
(D-1)-simplex seen from one of the two D-simplices (interaction
vertices) sharing it, so that the choice of kinetic functions
$\mathcal{K}$ specifies how the information and therefore the
geometric degrees of freedom corresponding to their D (D-2)-faces are
propagated from one vertex of interaction (fundamental spacetime
event) to another. We will detail this structure when describing the
Feynman graphs and the quantum dynamics of the theory. Before going on
to that, let us discuss briefly the classical dynamics of GFTs. What
we have is an almost ordinary field theory, in that we can rely on
a fixed background metric structure, given by the invariant
Killing-Cartan metric, and the usual splitting between kinetic
(quadratic) and interaction (higher order) term in the action, that
will later allow for a straightforward perturbative
expansion. However, the action is also {\it non-local} in that the
arguments of the D+1 fields in the interaction term are not all
simultaneously identified, but only pairwise; while this is certainly
a complication with respect to usual field theories in Minkowski
space, on the other hand this pairwise identification implies that,
even if the interaction is of order D+1, in terms of number of fields
involved, it is still quadratic in terms of the individual arguments
of the fields, which is in some sense a simplification of the usual
situation (e.g. with respect to renormalisation issues). The classical equations of motion following from the above
action are (schematically):
$$
\prod_j\int d\tilde{g}_{1j}\mathcal{K}(g_{1j}\tilde{g}_{1j}^{-1})\phi(\tilde{g}_{1j}) +
  \frac{\lambda}{D+1}\left(\prod_{j=1}^{D+1}\int dg_{2j}..dg_{D+1 j}\right)
  \phi(g_{2j})...\phi(g_{D+1 j})\,\mathcal{V}(\{g_{ij}\})=0
$$
and are therefore, assuming $\mathcal{K}$ to be a generic differential
operator, of integral-differential type. No detailed analysis of these
equations in any specific GFT model nor of their solutions has been
carried out to date, but work on this is in progress
\cite{etera-laurent-aristide}. The role and importance of these
equation from the point of view of the GFT per se are
obvious: they define the classical dynamics of the field theory, they
would allow the identification of classical background configurations
around which to expand in a semi-classical perturbation expansion,
etc. However, what is their meaning from the point of view of quantum
gravity, in light of the geometric interpretation of the field itself
as representing a (D-1)-simplex and of the GFT as a {\it local
  simplicial \lq 3rd quantisation\rq  of gravity}? The answer is simple
if striking: just as the Klein-Gordon equation represents at the same
time the classical equation of motion of a (free) scalar field theory
and the full quantum dynamics for the corresponding 1st quantised
(free) theory, in the same way these classical GFT equations encode
fully the quantum dynamics of the underlying (simplicial) canonical quantum gravity
theory, i.e. the one in which quantum gravity wave functions are
constructed, as for example in loop quantum gravity, by imposing the
quantum version of the classical Hamiltonian and diffeomorphism
constraints. This
means that solving the above equations amounts to identifying
non-trivial quantum gravity wave functions satisfying {\bf all} the
quantum gravity constraints, an important and still unachieved goal of
canonical quantum gravity. We will come back to this canonical
interpretation of GFTs after having introduced the perturbative
expansion of their partition function in Feynman diagrams.

Another issue that still needs a careful investigation is that of the
classical symmetries of the above action and of the resulting
equations. It should be obvious, given the previous discussion, that
the symmetries should correspond to the full set of symmetries of the
simplicial theory one is quantising. Some of these symmetries, holding
regardless of the specific choice of kinetic and interaction
operators, are the above-mentioned \lq closure\rq symmetry imposed on
each field: $\phi(g_i)=\phi(g_i g)$, $\forall g\in G$, encoded in the
symmetry property of the kinetic and vertex operators:
$\mathcal{K}(g_i\tilde{g}_i^{-1})=\mathcal{K}(g
g_i\tilde{g}_i^{-1}g')$,
$\mathcal{V}(g_{ij}\tilde{g}_{ji}^{-1})=\mathcal{V}(g_i
g_{ij}\tilde{g}_{ji}^{-1}g_j)$,  and the global symmetry of the action
under : $\phi(g_i)\rightarrow \phi(g g_i) \,\,\forall g\in G$. The
presence of additional symmetries depends on the specific choice of
kinetic and vertex operators, as said, as they would corresponds to
specific symmetries of the classical discrete theories being
quantised. The identification of such GFT analogues of the classical
symmetries of specific discrete theories on spacetime is no easy task,
and it is known already in the simpler example of GFT
formulations of BF theories that there exist characteristic symmetries
of the BF theories, i.e. so-called translation or topological
symmetries, that can be correctly identified at the level of the GFT
Feynman amplitudes, as identities between amplitudes for different
graphs, but do not correspond to the above obvious symmetries of the
GFT action \cite{laurentdiffeo}.

\
\

Let us now turn to the quantum dynamics of group field theories. Most
of the work up to now has focused on the perturbative aspects of this
quantum dynamics, and basically the only thing that has been explored
to some extent is the expansion in Feynman diagrams of the GFT
partition function and the properties of the resulting Feynman
amplitudes.
This expansion is given by:

$$ Z\,=\,\int
\mathcal{D}\phi\,e^{-S[\phi]}\,=\,\sum_{\Gamma}\,\frac{\lambda^N}{sym[\Gamma]}\,Z(\Gamma),
$$
where $N$ is the number of interaction vertices in the Feynman graph
$\Gamma$, $sym[\Gamma]$ is a symmetry factor for the graph and
$Z(\Gamma)$ the corresponding Feynman amplitude.
The Feynman amplitudes can be constructed easily after identification
of the propagator, given by the inverse of the kinetic term in the
action, and the vertex amplitude of the theory; each edge of the
Feynman graph is made of $D$ strands,
one for each argument of the field, and each one is then re-routed at the
interaction vertex, with the combinatorial structure of an
$D$-simplex, following the pairing of field arguments in the vertex operator. Diagrammatically:
\begin{figure}[here]
\setlength{\unitlength}{1cm}
\begin{minipage}[b]{5.9cm}
\includegraphics[width=5.5cm, height=3cm]{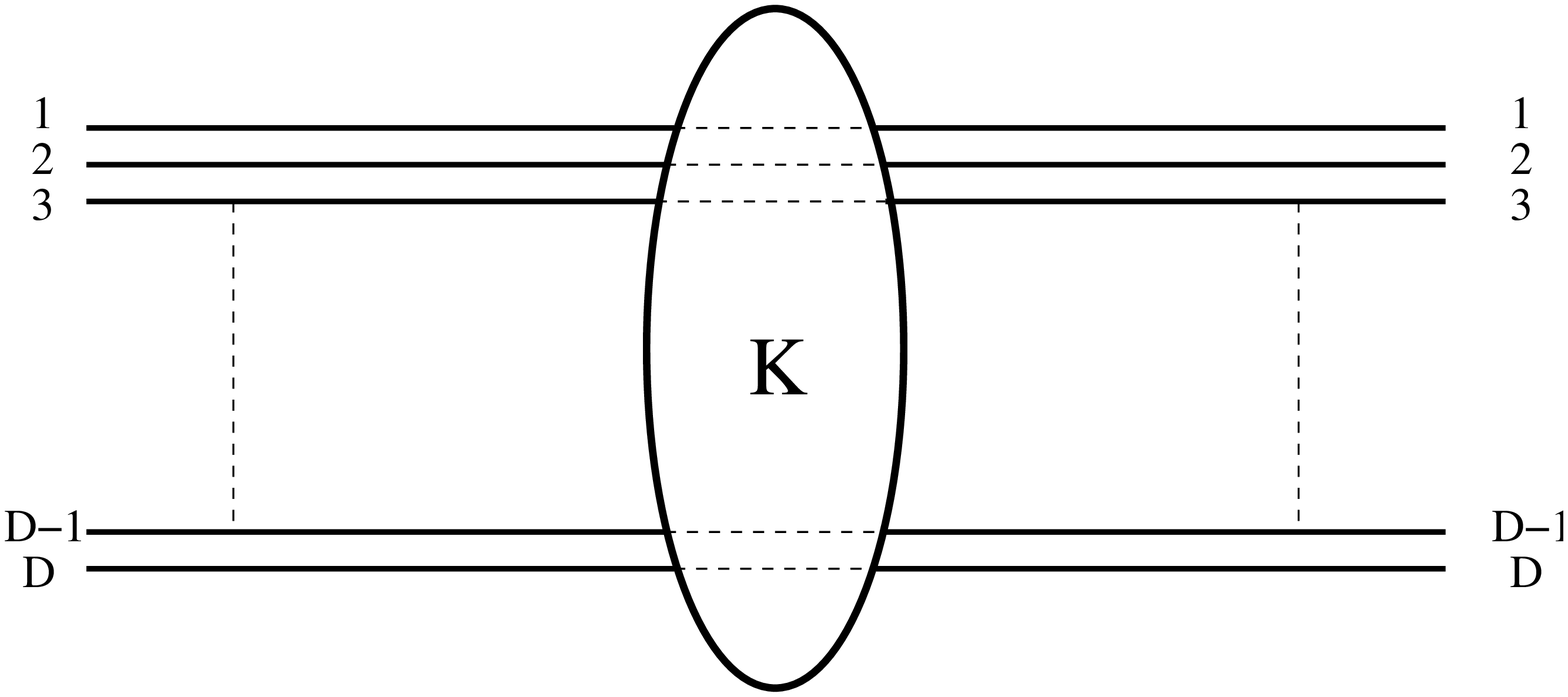}
\end{minipage}
\hspace{3cm}
\begin{minipage}[b]{5.2cm}
\includegraphics[width=5.5cm, height=4cm]{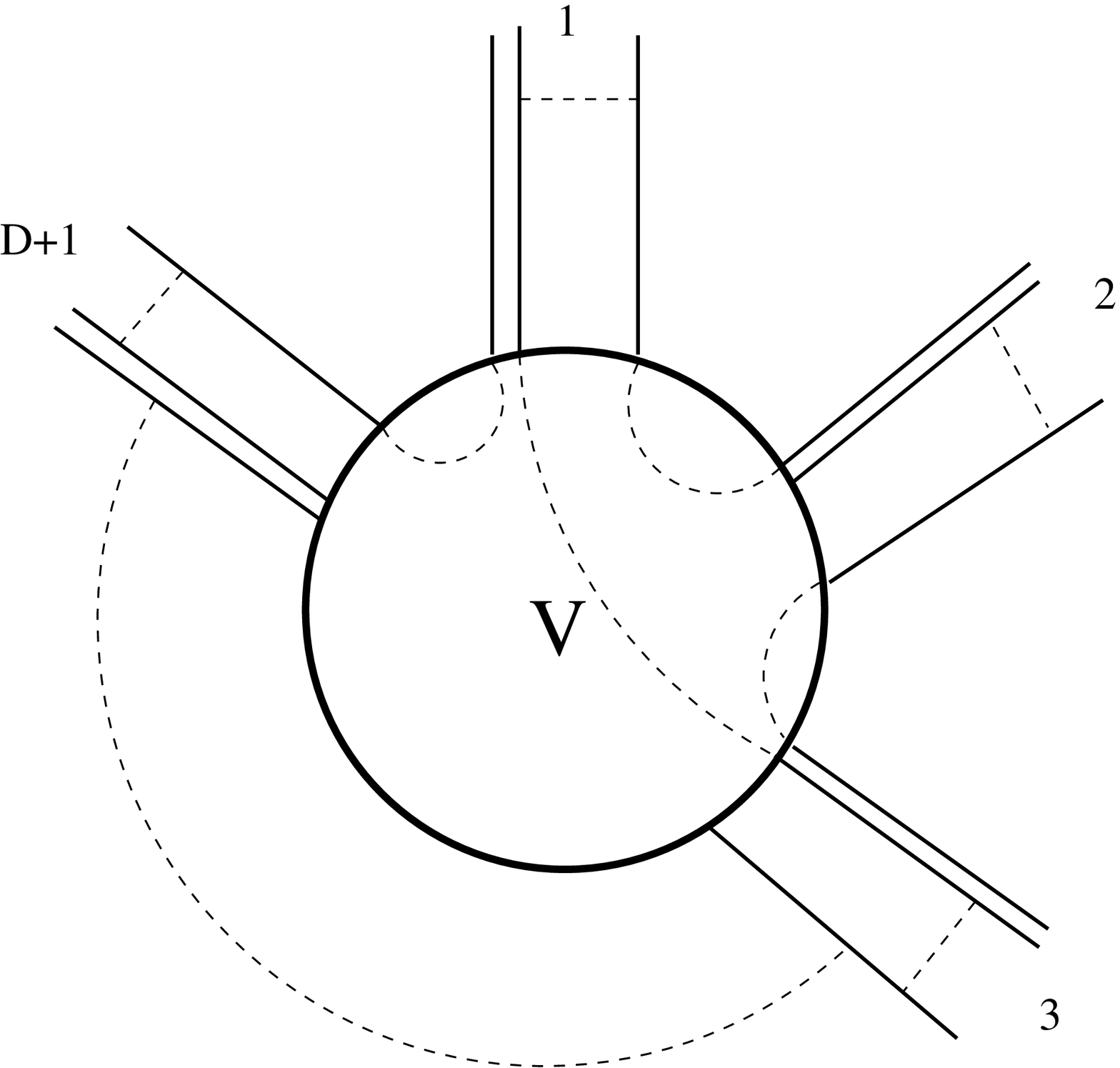}
\end{minipage}
\end{figure}

Each strand in an edge of the Feynman graph goes through several
vertices, coming back where it started, for closed Feynman
graphs, and therefore identifies a 2-cell (for open graphs, it may end
up on the boundary, identifying an open 2-cell). Each Feynman graph
$\Gamma$ is then a collection of 2-cells (faces), edges and vertices, i.e. a
2-complex, that, because of the chosen combinatorics for the arguments
of the field in the action, is topologically dual to a D-dimensional simplicial
complex \cite{DP-F-K-R,DP-P}. Clearly, the resulting complexes/triangulations can have
arbitrary topology, each corresponding to a particular {\it scattering
  process} of the fundamental building blocks of space,
i.e. (D-1)-simplices.

Even if every 2-complex arising as a GFT Feynman graph can be
understood as topologically dual to a D-dimensional triangulation,
this would not necessarily be a simplicial {\it manifold}. The data in
the GFT Feynman graphs do not constrain the neighbourhoods of simplices
of dimensions from (D-3) downwards to be spheres. This implies that in the
general case, the resulting simplicial complex, obtained by gluing
D-simplices along their (D-1)-faces, would correspond to a {\it
  pseudo-manifold}, i.e. to a manifold with {\it conical singularities}
\cite{DP-F-K-R, DP-P, cranematter}. A precise set of conditions under which
the GFT Feynman graphs correspond to manifolds is identified and
discussed at length in \cite{DP-P}, both at the level of simplicial
complexes and of the corresponding dual 2-complexes, in 2, 3 and 4
dimensions. All the relevant conditions can be checked algorithmically
on any given Feynman graph. It is not clear, at present, whether and
how one can construct suitably constrained GFT models satisfying these
conditions, i.e. that would generate only
manifold-like complexes in their Feynman expansion.

Each strand carries a field variable, i.e. a group element in
configuration space or a representation label in momentum
space. Therefore in momentum space each Feynman graph is given by a
spin foam (a 2-complex with faces labelled by representation variables), and each Feynman amplitude (a complex function of the
representation labels, obtained by contracting vertex amplitudes with
propagator functions) by a spin foam model (see chapter by Perez):

$$
Z(\Gamma)=\sum_{J_f}\prod_f A(J_f) \prod_e A_e(J_{f\mid e}) \prod_v A_v(J_{f\mid v}).
$$
As in all spin foam models, the representation variables have a
geometric interpretation (edge lengths, areas, etc) (see \cite{review,alexreview}) and so each of these Feynman amplitudes corresponds to a
definition of a sum-over-histories for discrete quantum gravity on the
specific triangulation dual to the Feynman graph, although the quantum
amplitudes for each geometric configuration are not necessarily given
by the exponential of a discrete gravity action.  For more on the
quantum geometry behind spin foam models (geometric meaning of the
variables, form and properties of the amplitudes, etc, we refer to the
literature \cite{review,alexreview, causal}.

One can show that the inverse is also true: any local spin foam model can
be obtained from a GFT perturbative expansion
\cite{carlomike,laurentgft}. This is a particularly important
result, since it implies that the GFT approach {\it subsumes} the spin
foam in its perturbative aspects, while at the same time going beyond
it, since there is of course much more in a QFT than its
perturbative expansion. The sum over Feynman graphs gives then a sum
over spin foams (histories of the spin networks on the boundary in any scattering process considered),
and equivalently a sum over triangulations, augmented by a sum over
algebraic data (group elements or representations) with a geometric
interpretation, assigned to each triangulation. This perturbative expansion of the
partition function also allows for a perturbative evaluation
of expectation values of GFT observables, as in ordinary QFT. These are given
\cite{laurentgft} by gauge invariant combinations of the basic field
operators, that can be constructed in momentum space using spin networks according to the formula:
$$ O_{\Psi=(\gamma, j_e,i_v)}(\phi)=\left(\prod_{(ij)}dg_{ij}dg_{ji}\right) \Psi_{(\gamma, j_e,i_v)}(g_{ij}g_{ji}^{-1})\prod_i \phi(g_{ij}),$$ where  $\Psi_{(\gamma, j_e,i_v)}(g)$ identifies a spin network functional (see chapter by Thiemann) for the spin network labelled by a graph $\gamma$ with representations $j_e$ associated to its edges and intertwiners $i_v$ associated to its vertices, and $g_{ij}$ are group elements associated to the edges $(ij)$ of $\gamma$ that meet at the vertex $i$.

In particular,
the transition amplitude (probability amplitude for a certain
scattering process) between certain boundary data represented by two
spin networks, of arbitrary combinatorial complexity, can be expressed as the expectation value of the field
operators having the same combinatorial structure of the two spin
networks \cite{laurentgft}.

$$
\langle \Psi_1\mid\Psi_2\rangle = \int \mathcal{D}\phi\,O_{\Psi_1}\,O_{\Psi_2}\,e^{-S(\phi)} = \sum_{\Gamma/\partial\Gamma=\gamma_{\Psi_1}\cup\gamma_{\Psi_2}}\,\frac{\lambda^N}{sym[\Gamma]}\,Z(\Gamma)
$$
where the sum involves only 2-complexes (spin foams) with boundary given by the two spin networks chosen.
\
\
The above perturbative expansion involves therefore two very different
types of sums: one is the sum over geometric data (group elements or
representations of $G$) entering the definition of the Feynman
amplitudes as the GFT analogue of the integral over momenta or
positions of usual QFT; the other is the overall sum over Feynman
graphs; the last sum includes a sum over all triangulations for a
given topology and a sum over all topologies\footnotemark\footnotetext{Obviously, given
  that no algorithmic procedure exists that allows to distinguish
  topologies in $D\geq 3$, we cannot partition this sum into the two
  sub-sums mentioned; however, we are assured that all topologies are
  present in the full sum because, as already mentioned, any simplicial complex of any topology is obtained by
  an appropriate gluing and face identification of fundamental
  simplicial building blocks, and that all such gluings and
  identifications are present by construction in the GFT Feynman
  expansion.}.
Both sums are potentially divergent. First of all the naive definition
of the Feynman amplitudes implies a certain degree of redundancy,
depending on the details of the kinetic and vertex operators chosen,
resulting from the symmetries of the defining GFT. This
means that a proper gauge fixing of these symmetries, especially those
whose group is non-compact, is needed to avoid the
associated divergences \cite{laurentdiffeo}. Even after gauge fixing, the sum over geometric data has a
potential divergence for every \lq bubble\rq  of the GFT Feynman
diagram, i.e. for every closed surface of it identified by a
collection of 2-cells. This is the GFT analogue of loop divergences of
usual QFT. Of course, whether the GFT amplitudes are divergent or not
depends on the specific model. For example, while the most natural
definition of the group field theory for the Barrett-Crane spin foam
model \cite{DP-F-K-R}, presents indeed bubble divergences, a simple
modification of it \cite{P-R}, producing a different version of the
same model, possesses {\it finite} Feynman amplitudes, i.e. it is {\it
  perturbatively finite} without the need for any regularization. In
general however a regularization and perturbative renormalisation
procedure would be needed to make sense of the sums involved in the
definition of a given GFT. No systematic study of renormalisation of
GFTs has been carried out to date. The importance and need of this
study is obvious, given that GFTs are presently {\it defined} by their
perturbative expansion and that only a proof of renormalisability of
such model would ensure that such perturbative expansions make
physical sense.
The sum over Feynman graphs, on the other hand, is most certainly
divergent. This is not surprising both from the quantum gravity point
of view and the QFT one. As said, the sum over Feynman graphs gives a
sum over {\it all} triangulations for {\it all} topologies, each
weighted by a (discrete) quantum gravity sum-over-histories. That such
a sum can be defined constructively thanks to the simplicial and QFT
setting is already quite an achievement, and to ask for it to be
finite would be really too much! Also, from the strictly QFT
perspective, i.e. leaving aside the quantum gravity interpretation of
the theory and of its perturbative expansion, it is to be expected
that the expansion in Feynman graphs of a QFT would produce at most an
asymptotic series and not a convergent one. This is the case for all
the interesting QFTs we know of. What makes the usual QFT perturbative expansion
useful in spite of its divergence is the simple fact that it has a
clear physical meaning, in particular the fact that one knows what it
means to compute a transition amplitude up to a given order. In the
GFT case this means, among other things, providing a clear physical
interpretation to the coupling constant $\lambda$. This can be done,
actually, in more than one way. First of all, defining
$\alpha=\lambda^{\frac{1}{D-1}}$ and redefining
$\tilde{\phi}=\alpha\phi$, we can recast the GFT action in the form
$S_{\lambda}[\phi]=\frac{1}{\alpha^2}S_{\lambda=1}[\tilde{\phi}]$. One
can then perform a loop expansion of the GFT partition function, that
is an expansion in the parameter $\alpha$, instead of a perturbative
expansion in the coupling constant. This gives, for a generic
transition amplitude between two boundary states $\Psi_1$ and
$\Psi_2$:
$\langle\Psi_1\mid\Psi_2\rangle_\alpha=\frac{1}{\alpha^2}\sum_{i=0}^{\infty}\alpha^{2
  i} \langle\Psi_1\mid\Psi_2\rangle_i$, where
$\langle\Psi_1\mid\Psi_2\rangle_i$ is a sum over Feynman graphs with
$i$ loops. The point here is to realise that adding a loop to a given
Feynman diagram is equivalent \cite{laurentgft} to adding a handle to
the simplicial complex dual to it. This means that the parameter
$\alpha=\lambda^{\frac{1}{D-1}}$ governs the strength of topology
changing processes in the GFT perturbative expansion. This
interpretation can also be confirmed by analysing the Schwinger-Dyson
equations for a generic GFT \cite{laurentgft}.
A different perspective on the physical meaning of $\lambda$ is
obtained by noticing that $\lambda$ weights somehow the \lq size\rq of
the spacetimes emerging in the GFT perturbative expansion, assuming
that the number of D-simplices is a measure
of the D-volume of spacetime. Then one could define $\lambda =
e^{i\Lambda}$, interpret the Feynman amplitude as defining the
exponential of an (effective) action for gravity
$Z(\Gamma)=e^{iS(\Gamma)}$ and thus rewrite heuristically the GFT
partition function as: $Z = \sum_\Gamma \frac{1}{sym(\Gamma)} e^{i (
  S(\Gamma) + \Lambda V(\Gamma))}$. Now, assuming for $S(\Gamma)$ a
form given by, say, the Regge action for pure gravity with no
cosmological constant on a triangulation (dual to $\Gamma$) with fixed
edge lengths, then $\Lambda$ would play the role of a cosmological
constant contribution to the gravity action. Indeed this would be
exactly the expression for a dynamical triangulations model
\cite{renate} with $\Lambda$ being the bare cosmological
constant. While this is just a very heuristic argument and there is no
easy way to make it rigorous for a general GFT, it can in fact be made
rigorous by considering a tensor model \cite{Ambjorn,DP-P}, a
special case of the GFT formalism. This is obtained from the GFT for
BF theories roughly by constraining the representations of $G$ used to
be all equal to a given one with dimension $N$; we will give more
details on both tensor models and GFTs for BF theories in the
following. For tensor models, the Feynman expansion of the partition function is given by \cite{DP-P}: $$ Z=\sum_{T}
\frac{1}{\textnormal{sym}(T)}\lambda^{n_D} N^{n_{D-2}}. $$ By
redefining
$\lambda=e^{-(\Lambda \,V_{D}+\frac{D(D+1)}{2}\frac{\arccos{1/D}}{16\pi
    G}\,V_{D-2})}$ and
$N=e^{\frac{V_{D-2})}{8G}}$, where $G$ is the bare
Newton constant, $V_D$ is the volume of an equilateral
D-simplex with edge lengths equal to $a$, and $V_{D-2}$ is
the volume of an equilateral (D-2)-simplex, the expression above turns
into that defining the dynamical triangulations approach
\cite{renate}, if the triangulations are restricted to those defining
manifolds as opposed to pseudo-manifolds. Here, as anticipated,
$\Lambda$ plays the role of the bare cosmological constant. As said,
it is not obvious whether and how such interpretation of the GFT
coupling constant $\lambda$ extends to the general GFT case, and more
work on this is needed. However, we want to stress that, on the one
hand, the two proposed interpretations for $\lambda$ are compatible with each other, and that finding a clear link between the two would mean
linking the value of the bare, and then of the renormalised,
cosmological constant to the presence of spatial topology change. This
would be the realisation in a more rigorous context of an old idea
\cite{banks, coleman} that was indeed among the first motivations for
developing a \lq 3rd quantisation\rq  formalism for quantum gravity.

\
\

Before going on to discuss some specific GFT models, we would like to
remark once more on the connection between GFT and canonical quantum
gravity, and to point out to one important recent proposal concerning
this relation.
As already mentioned, the classical GFT equations of motion are
interpreted as encoding the full quantum dynamics of the corresponding
1st quantised theory; this is a simplicial quantum gravity theory
whose kinematical quantum states are labelled by D-valent spin
networks for the group $G$. If the spin foams describing the histories
of such states were to be restricted to those corresponding to
triangulation of trivial spacetime topology, then a sum over such
histories weighted by an appropriate amplitude would be a possible
definition of the canonical inner product for a simplicial version of
loop quantum gravity based on the group $G$, if the 2-point function
so defined is positive semi-definite as it should. In other words such
sum-over-histories would define implicitly a projection onto physical
quantum states of gravity, thus fully encoding the dynamics of the
theory usually given in terms of an Hamiltonian constraint operator
(see chapters by Thiemann and Perez). Now, the GFT formalism allows a
natural definition of such sum, that involves, as expected, only the
classical information encoded in the GFT action. In fact, one could
consider the restriction of the GFT perturbative expansion given above
to {\it tree level}, for given boundary spin network
observables,\cite{laurentgft}:

$$
\langle \Psi_1\mid\Psi_2\rangle = \sum_{\Gamma_{\textnormal{tree}}/\partial\Gamma=\gamma_{\Psi_1}\cup\gamma_{\Psi_2}}\,\frac{\lambda^N}{sym[\Gamma]}\,Z(\Gamma).
$$

This is the GFT definition of the canonical inner product, i.e. of the
matrix elements of the projector operator \cite{carlomike}, {\it for
  Feynman amplitudes that are real and positive}, as for example those
of the BF or Barrett-Crane models. The definition is well posed,
because at tree level every single amplitude $Z(\Gamma)$ is finite
whatever the model considered due to the absence of infinite summation
(unless it presents divergences at specific values of the
configuration/momentum variables). Moreover, it possesses all the
properties one expects from a canonical inner product: 1) it involves
a sum over Feynman graphs, and therefore triangulations, with the
cylindrical topology $S^{D-1}\times [0,1]$, for closed spin networks
$\Psi_i$ associated with the two boundaries, as it is easy to verify;
2) it is real and positive, as said, but not strictly positive; it has
a non-trivial kernel that can be shown \cite{laurentgft} to include
all solutions of the classical GFT equations of motion, as
expected. This means that the physical Hilbert space for canonical
spin network states can be constructed, using the GNS construction,
from the kinematical Hilbert space of all spin network states by
modding out those states belonging to the kernel of the above inner
product.

On the one hand this represent a concrete (and the first) proposal for
completing the definition of a loop formulation of quantum gravity,
that can now be put to test, so the possible end of a long quest
started more than 15 years ago, and thus a proof of the
usefulness of GFT ideas and techniques if only with respect to the
canonical/Hamiltonian quantum gravity program. On the other hand, and,
in our opinion, most importantly, it makes clear that the GFT
formalism contains and therefore can achieve much more than any
canonical quantum theory of gravity, given that the last is fully
contained in the very small subset of the former corresponding to the
\lq classical\rq   level only. Within the GFT formalism, we can go
beyond this investigating, for example, the full
quantum structure, topology change, quantum gravity 2-point functions (in \lq
superspace\rq) that are {\bf not} the canonical inner product
\cite{generalised}, etc.

\section{Some group field theory models}

\subsection{Pure gravity}
Let us now discuss some specific GFT models. The easiest
example is the straightforward generalisation of matrix models for 2d
quantum gravity to a GFT \cite{eac}, given by the action:
\begin{equation}
S[\phi]=\int_G dg_1dg_2 \frac{1}{2}\phi(g_1,g_2)\phi(g_1,g_2) +
\frac{\lambda}{3!}\int dg_1dg_2dg_3\phi(g_1,g_2)\phi(g_1,g_3)\phi(g_2,g_3)
\end{equation}
where $G$ is a generic compact group, say $SU(2)$, and the symmetries
mentioned above are imposed on the field $\phi$ implying, in this
case: $\phi(g_1,g_2)=\tilde{\phi}(g_1g_2^{-1})$. The relation with matrix
models is apparent in momentum space, expanding the field in
representations $j$ of $G$ to give:

\begin{equation}
S[\phi]=\sum_j \textit{dim}(j) \left( \frac{1}{2}tr{(\tilde{\phi}_j^2)} +\frac{\lambda}{3!}tr(\tilde{\phi}_j^3)\right)
\end{equation}

where the field modes $\tilde{\phi}_j$ are indeed matrices with
dimension $\textit{dim}(j)$, so that the action is given by a sum of
matrix models actions for increasing dimensions. Alternatively, one
can see the above as the action for a single matrix model in which the
dimension of the matrices has been turned from a parameter into a
dynamical variable. The Feynman amplitudes for such a theory are given
by $Z(\Gamma)=\sum_j \textit{dim}(j)^{2-2 g(\Gamma)}$ and shows that
the GFT above gives a quantisation of BF theory (with gauge group $G$)
on a closed triangulated surface, dual to $\Gamma$, of genus
$g(\Gamma)$, augmented by a sum over all such surfaces defined as a
Feynman expansion\cite{eac}. A quantisation of 2d gravity along the same lines
would use an $U(1)$ connection (and therefore $G=U(1)$) and a
restriction on the representations summed over to  single one,
together with additional data encoding bundle information in the GFT
action, see \cite{iocarlosimone}.

The extension to higher dimensions can proceed in two ways. The first
attempt at this was done in \cite{Ambjorn} where the first \lq tensor
model\rq was introduced, with action:

\begin{equation}
S[\phi]=\sum_{\alpha_{i}}\left(\frac{1}{2}\phi_{\alpha_1\alpha_2\alpha_3}\phi_{\alpha_1\alpha_2\alpha_3}
+\frac{\lambda}{4!}\phi_{\alpha_1\alpha_2\alpha_3}\phi_{\alpha_3\alpha_4\alpha_5}\phi_{\alpha_5\alpha_2\alpha_6}\phi_{\alpha_6\alpha_4\alpha_1}\right)
\label{eq:tensor},
\end{equation}

for a $N\times N\times N$ tensor $\phi$.

This model generates 3-dimensional simplicial complexes in its Feynman
expansion, that include \cite{Ambjorn, DP-P} both manifold- and
pseudo-manifold-like configurations. It is not however a GFT until
group structure is added. This is done again as a straightforward
generalisation of the above model for the 2d case. In fact the
following choice of kinetic and vertex term in the generic action \ref{eq:action}:

$$
\mathcal{K}(g_i,\tilde{g}_i) = \int_G dg \prod_i
\delta(g_i\tilde{g}_i^{-1}g),\;\;\;\;\;\;\mathcal{V}(g_{ij},g_{ji}) =
\prod_i\int_G dg_i \prod_{i<j}\delta(g_i g_{ij}g_{ji}^{-1}g_j^{-1}),
$$
where the integrals impose the gauge invariance under the action of
$G$, gives the GFT quantisation of BF theories, for gauge group $G$,
in any dimension \cite{boulatov, ooguri}. In particular, in 3 dimensions, the choice of
$G=SO(3)$ or $G=SO(2,1)$ provides a quantisation of 3D gravity
in the Euclidean and Minkowskian signatures, respectively, and the
so-called Ponzano-Regge spin foam model, as evaluation of the GFT
Feynman amplitudes (see also \cite{iogft} for more details on this
GFT, first proposed by Boulatov \cite{boulatov}), while the choice of
the quantum group $SU(2)_q$ gives the Turaev-Viro topological
invariant.
For the $SU(2)$ case, the action is then:
\begin{eqnarray}
S[\phi]=\prod_i \int_{SU(2)} \phi(g_1,g_2,g_3)\phi(g_1,g_2,g_3)+
\frac{\lambda}{4!}\prod_{i=1}^{6}\int_{SU(2)}dg_i \;\phi(g_1,g_2,g_3)\phi(g_3,g_4,g_5)\phi(g_5,g_2,g_6)\phi(g_6,g_4,g_1).
\;\;\;\;\end{eqnarray}

Lots is known about the last model (see chapter by
Freidel), including the appropriate gauge fixing procedure of its
Feynman amplitudes, the coupling of matter fields, etc. Here, we
mention only one result that is of interest for the general issue of
GFT renormalisation. This is the proof \cite{laurentborel} that a
simple modification of the GFT above gives a model whose
perturbative expansion is Borel summable. The modification amounts to
adding another vertex term of the same order to the original one, to give:

\begin{eqnarray*}
+\frac{\lambda}{4!}\prod_{i=1}^{6}\int_{SU(2)}dg_i\left[
\phi(g_1,g_2,g_3)\phi(g_3,g_4,g_5)\phi(g_5,g_2,g_6)\phi(g_6,g_4,g_1)+ \,\delta\, \phi(g_1,g_2,g_3)\phi(g_3,g_4,g_5)\phi(g_4,g_2,g_6)\phi(g_6,g_5,g_1)\right].
\end{eqnarray*}

The new term corresponds simply to a slightly different
recoupling of the group/representation variables at each vertex of
interaction, geometrically to the only other possible way of gluing 4
triangles to form a closed surface. This Borel-summability is
interesting for more than one reason: 1) it shows that it {\it is} possible
to control the sum over triangulations of all topologies appearing in
the GFT perturbative expansion, despite the factorial growth in their
number with the number of 3-simplices forming them; whether the sum is
summable or not, then, depends clearly only on the amplitude weighting
them; 2) even if the above modification of the Boulatov GFT model has no clear physical interpretation
yet from the quantum gravity point of view, e.g. in terms of gravity
coupled to some sort of matter, it is indeed a very mild modification,
and most importantly one that one would expect to be forced upon us by
renormalisation group-type of argument, that usually require us to
include in the action of our field theory all possible terms that are
compatible with the symmetries.

Still in the 3D case, it is easy to check that a GFT action with the same
combinatorial structure of the Boulatov model, but for a real field defined
on the homogeneous space $SO(3)/SO(2)\simeq S^2$ (in turn obtained
choosing $G=SO(3)$ and then averaging over the subgroup $SO(2)$ in
each argument of the field \cite{laurentborel}), and with the
invariance under the global $SO(3)$ action in each field having been
dropped, gives a generalisation of the tensor model \ref{eq:tensor}
analogous to the GFT generalisation of matrix models previously
described. In fact the action for this model, in momentum space, is:

$$
S[\phi]=\sum_{j_i,\alpha_{i}}\left(\frac{1}{2}\phi^{j_1j_2j_3}_{\alpha_1\alpha_2\alpha_3}\phi^{j_1j_2j_3}_{\alpha_1\alpha_2\alpha_3}
+\frac{\lambda}{4!}\sum_{j_i,\alpha_i}\phi^{j_1j_2j_3}_{\alpha_1\alpha_2\alpha_3}\phi^{j_3j_4j_5}_{\alpha_3\alpha_4\alpha_5}\phi^{j_5j_2j_6}_{\alpha_5\alpha_2\alpha_6}\phi^{j_6j_4j_1}_{\alpha_6\alpha_4\alpha_1}\right)
$$ where the indices $\alpha_i$ run over a basis of vectors in the
representation space $j_i$,
and its partition function is:
$Z=\sum_{\Gamma}\frac{(-\lambda)^{n_\Gamma}}{\textnormal{sym}(\Gamma)}\sum_{j_f}\prod_f
(2j_f+1)$, with $f$ being th faces of the 2-complex/Feynman graph,
which is divergent and has to be regularised. There are three ways of
doing it, and, by doing this, reducing the partition function to that
of a tensor model: 1) simply dropping the sum over the
representations $j_f$ by fixing them to equal a given $J$; 2) placing
a cut-off on the sum by restricting $j_i<N$, obtaining
$Z=\sum_{\Gamma}\frac{(-\lambda)^{n_v(\Gamma)}}{\textnormal{sym}(\Gamma)}[(N+1)^2]^{n_f(\Gamma)}$;
    3) equivalently, but more elegantly, by defining the model not on
    $S^2$ but on the non-commutative 2-sphere $S^2_N$, which also
    carries a representation of $SU(2)$ but implies a bounded
    decomposition in spherical harmonics (labelled by $j<N$), thus
    giving the same result for the partition function. We recognise in
    the above result the partition function for the tensor model
    \ref{eq:tensor} and for a dynamical triangulations model \cite{renate}.

Let us now discuss the situation in the 4D case. Here GFT model building has followed the development of spin foam models
for 4D quantum gravity. The guiding idea in the spin foam setting (see
chapter by Perez) has been the fact that {\it
  classical} gravity can be written \cite{Peldan, FKP} as a
constrained version of a BF theory for the Lorentz group; the
construction of the appropriate spin foam model then required
\cite{BC} first a description of simplicial geometry
in terms of bivectors (discretion of the $B$ field of the BF
theory) subject to the discrete analogue of the
continuum constraints \cite{BC}, and then finding the algebraic
translation of these at the quantum level as constraints on the
representations labelling the spin foam faces. The Barrett-Crane spin
foam models obtained as a result of this amount roughly to a restriction of
spin foam models for BF theories to involve so-called {\it simple}
representations only of the Lorentz group ($SO(4)$ or $SO(3,1)$, or
the corresponding double covering,
depending on the signature wanted for spacetime) \cite{BC, alexreview,
  review}. The same restriction can be imposed at the GFT level,
starting from the GFT describing 4D BF theory, by projecting down the
arguments of the field from $G=SO(4)$ ($SO(3,1)$) to the homogeneous
space $SO(4)/SO(3)\simeq S^3$ ($SO(3,1)/SO(3)$ or
$SO(3,1)/SO(2,1)$ in the Lorentzian case), exploiting the fact that
only simple representations of $G$ appear in the harmonic
decomposition of functions on these spaces. The GFT action is then
defined \cite{DP-F-K-R}
(we deal explicitly only with the Riemannian case for simplicity) as:

\begin{eqnarray}
 \lefteqn{S[\phi]=\frac{1}{2}\left(\prod_i \int_{SO(4)} dg_i\right)
 P_gP_h\phi(g_1,g_2,g_3,g_4)P_gP_h\phi(g_1,g_2,g_3,g_4) +}\nonumber \\
&+&\frac{\lambda}{5!}\left(\prod_{i=1}^{10}\int_{SO(4)}dg_i\right)
 \left[
 P_gP_h\phi(g_1,g_2,g_3,g_4)P_gP_h\phi(g_4,g_5,g_6,g_7)P_gP_h\phi(g_7,g_8,g_3,g_9)\right.
 \nonumber \\ &\left.\right.& \left.\hspace{4cm}P_gP_h\phi(g_9,g_5,g_2,g_{10})P_gP_h\phi(g_{10},g_8,g_6,g_1)\right]\;\;\end{eqnarray}
where the projection $P_h\phi(g_i)=\prod_i\int_{SO(3)}dh_i \,\phi(g_i
h_i)$ from the group to the homogeneous space imposes the wanted
constraints on the representations, in momentum space, and the
projection $P_g\phi(g_i)=\int_{SO(4)} dg \,\phi(g_i g)$ ensures that
gauge invariance is maintained.

Different variations of this model can be constructed \cite{P-R,
  thesis} by inserting the two projectors $P_h$ and $P_g$ in the
  action in different combinations. However, the symmetries of the
  resulting models remain the same, and the corresponding Feynman
  amplitudes have the structure:

\begin{equation}
Z(\Gamma)=\sum_{J_f}\prod_f \textnormal{dim}(J_f) \prod_e A_e(J_{f\mid e}) \prod_v \mathcal{V}_{BC}(J_{f\mid v}),
\end{equation}
where $\textnormal{dim}(J_f)$ is the measure for the representation
$J_f$, labelling the faces of the 2-complex/Feynman graph, entering
the harmonic decomposition of the delta function on the group, and the
function $\mathcal{V}_{BC}(J_{f\mid v})$, depending on the ten
representations labelling the ten faces of $\Gamma$ incident to the
same vertex $v$ is the so-called Barrett-Crane vertex
\cite{BC,review,alexreview}; the exact form of the edge amplitude $A_e$, on
the other hand, depend on the version of the specific variation of the
above action being considered.

The above Feynman amplitudes, modulo the exact form of the edge
amplitudes, can be obtained and justified in various ways,
e.g. starting from a discretisation of classical BF theory and a
subsequent imposition of the constraints \cite{review,alexreview}, and
in particular there is a good consensus on the validity of the
Barrett-Crane construction for the vertex amplitude. As a hint
that at least {\it some} of the properties of gravity we want are
correctly encoded in the above Feynman amplitudes, we cite the fact
that there exists at least a sector of the geometric configurations
summed over where an explicit connection with classical discrete
gravity can be exhibited. In fact \cite{ruthjohn}, for non-degenerate configurations
(i.e. those corresponding to non-degenerate simplicial geometries) the
asymptotic limit of the Barrett-Crane amplitude $\mathcal{V}_{BC}(J)$
is proportional to the cosine of the Regge action for simplicial
gravity, i.e. the correct discretion of the Einstein-Hilbert
action for continuum General Relativity. It is therefore very close to
the form one would have expected for the amplitudes in a
sum-over-histories formulation of quantum gravity on a given
simplicial manifold.

All the above models (for BF theory and quantum gravity) in both 2, 3 and 4 dimensions
share the following general properties: 1) their Feynman amplitudes
are real; 2) they do not depend on the
orientation chosen for the complexes, or to put it differently, no
unique orientation for the (various elements of
the) triangulation dual to any Feynman graph can be reconstructed from the amplitude associated to it; 3) in quantum gravity models, the
asymptotic limit of the vertex amplitude gives (in the non-degenerate
sector) the {\it cosine} of the Regge action instead of the
exponential of it. These properties suggest the interpretation
\cite{generalised} of the corresponding models as defining the quantum
gravity analogue of the Hadamard 2-point function for a relativistic
particle (anti-commutator of field operators in QFT), or, more
appropriately in our field theory context, as defining a QFT picture
for quantum gravity with Hadamard propagators used everywhere in the
\lq internal lines\rq of the graphs, instead of the usual Feynman
propagators. Also, these are the wanted properties if our aim is
to define, through a GFT, the inner product of a canonical/Hamiltonian
theory of quantum gravity based on spin networks, as we have
discussed, i.e. we want to impose the dynamics encoded in a
Hamiltonian constraint using covariant methods. In other words, the Feynman amplitudes of these GFT
models would correspond not to a simplicial version of the path
integral formalism for quantum gravity, but to the symmetrized version
of the same over opposite spacetime orientations, that indeed gives a
path integral definition of solutions of the Hamiltonian constraint
operator of canonical quantum gravity \cite{halliwellhartle}. However,
there are several reasons why one may want to go beyond this type of
structure. We cite three of them. 1) From a field theoretic
perspective applied to quantum gravity, i.e. from the point of view of
a field theory on a simplicial superspace we are advocating here, the
most natural object one would expect a GFT to define with its 2-point
functions is not a canonical inner product, solution of the
Hamiltonian constraint, but a Green function for it. This is what
happens in ordinary QFT, for the free theory, and in the formal
context of continuum 3rd quantisation for quantum gravity. In the last
case, in fact, the Feynman amplitudes of the theory, in absence of
spatial topology change, correspond to the usual path integral
formulation of quantum gravity for given boundary wave functions, with
amplitude given by the exponential of the GR action \cite{giddstro},
which is a Green function for the Hamiltonian constraint, and not a
solution of the same, in each of its arguments.
2) The orientation of the GFT 2-complexes can be given, for Lorentzian
models, a {\it causal} interpretation \cite{causal,feynman}, if one
also interprets the vertices of the same as fundamental spacetime
events, and thus the orientation independence of the spin foam
amplitudes associated to the usual models implies that these models
define {\it a-causal} transition amplitudes for quantum gravity.
Again, this is consistent with the field-theoretic interpretation of
the same amplitudes given above and with the wish to construct the
canonical inner product, but it also suggest that one should be able
to construct other types of models defining {\it causal} quantum
gravity transition amplitudes \cite{causal,feynman} instead and thus
different types of GFTs. 3) If there is, as we have seen, a clear
interpretation from the perspective of canonical quantum gravity for
the amplitudes defined by the GFT in the case of trivial spacetime
topology (i.e. no spatial topology change), this occurs in a very
limited subsector of the theory, and no clear meaning can be given in
the same perspective to the GFT amplitudes for Feynman graphs beyond
the tree level, when spatial topology change is present, and thus a
canonical/Hamiltonian formalism is not well-defined. For all these
reasons one would like to be have a more general class of GFT models
that do depend on the orientation of the GFT Feynman graphs, that can
be interpreted consistently as analogues of causal transition
amplitudes of QFT, that are in more direct contact with usual path
integral formulations of (simplicial) gravity, and that reduce to the
above type of models (and corresponding amplitudes) when suitably
restricted. This was achieved in \cite{generalised}. Here a
generalised version of the GFT formalism was defined, for a field $\phi(g_i,s_i): (G\times\mathbb{R})^{\otimes
  4}\rightarrow\mathbb{C}$, with action (in the 4d Riemannian case):
\begin{eqnarray*}
\lefteqn{S_{gen}(\{\lambda\})=\sum_{\mu,\alpha}\frac{1}{4}\prod_{i=1}^{4}\int
dg_i\,\int_\mathbb{R}ds_i \left\{
\phi^{-\mu\alpha}(g_i,s_i)\left[\prod_i\left(
  -i\mu\alpha\partial_{s_i}+\nabla_{i}\right)\right]\phi^{\mu\alpha}(g_i,s_i)\right\} +} \nonumber \\ &+&\sum_{\mu}\sum_{\alpha_i}
  \frac{\lambda_{\{\alpha_i,\mu\}}}{5!}\prod_{i\neq
  j=1}^{5}\int_G dg_{ij}\int_\mathbb{R}ds_{ij}
  \left\{ P_h\phi^{\mu\alpha_1}(g_{1j},s_{1j})
P_h\phi^{\mu\alpha_2}(g_{2j},s_{2j})
........ \right.
\nonumber \\
&& \left. ........P_h\phi^{\mu\alpha_5}(g_{5j},s_{5j})
\prod\,\theta(\alpha_i s_{ij}+ \alpha_j s_{ji})K\left(g_{ij},
g_{ji};\mu(\alpha_i s_{ij} + \alpha_j
s_{ji})\right)\right\},\;\;\;\;
\end{eqnarray*}
 where: $g_i\in G=SO(4)$, $s_i\in\mathbb{R}$, $\mu=\pm 1$ and $\alpha_i = \pm 1$ are orientation
 data that allow to reconstruct the orientation of the Feynman graph
 from the {\bf complex} amplitude associated to it,
 $\phi^{+}(g_i,s_i)=\phi(g_1,s_1;...,g_4,s_4)$ and
 $\phi^{-}(g_i,s_i)=\phi^{\dagger}(g_i,s_i)$, $P_h$ is the projector
 imposing invariance under the $SO(3)$ subgroup, $\nabla$ is the
 D'Alambertian operator on the group $G$, $\theta(s)$ is the step
 function and $K(g,s)$ is the evolution kernel for a scalar particle
 on the group manifold $G$ with evolution parameter $s$. The field is
 assumed invariant under the diagonal action of $G$, as in the models discussed above. The form of the kinetic operator and the use of the
 evolution kernels with proper time parameter $s$, plus the
 restriction on its range (through the theta functions), just as in
 the definition of the Feynman propagator of a relativistic particle,
 in the vertex term, impose a non-trivial dependence on the
 orientation data in fully covariant way. The resulting Feynman
 amplitudes \cite{generalised} have all the properties wanted, being
 complex and orientation-dependent, and have the natural
 interpretation as analogues of Feynman transition amplitudes for
 quantum gravity \cite{generalised, feynman}. Also, when expressed in
 momentum space, but only for the variables $s_i$, whose conjugate
 variables are labelled suggestively as {\it variable masses} $m_i^2$,
 the amplitude for each vertex is given by:
\begin{eqnarray*}
\prod_{f_{ij}\mid v} \int_{0}^{+\infty}ds_f\,\theta(\alpha_i\alpha_j
s_f)\, K(\vartheta_{f_{ij}},\mu_v\alpha_{i}\alpha_{j}s_f)e^{-im_f^2s_f}=\prod_{f_{ij}\mid v}
\frac{1}{4\pi}\frac{1}{\sin\vartheta_{f_{ij}}}e^{i\mu_v
  \sqrt{1-\mu\alpha_{i}\alpha_{j}m^2_{f_{ij}}}\vartheta_{f_{ij}}},
\end{eqnarray*}
with $\sqrt{1-\mu\alpha_{i}\alpha_{j}m^2_{f_{ij}}}$ interpreted as
the area of the triangle dual to the face $f_{ij}$, and the angle
$\vartheta_{f_{ij}}$ measuring the holonomy around the portion of
the face dual to the same triangle inside the 4-simplex dual to
$v$; thus this amplitude is \cite{generalised} of the form of the
exponential of the Regge action in 1st order formalism, times an
appropriate measure factor. It remains to be proven that this form
still holds, in these variables, for the amplitude associated to
the whole Feynman graph \cite{IoTimGFT, GFTsimplicial}. Other
models in the same class, i.e. with the same type of fields and
the same symmetries, can be defined, and share similar properties
\cite{IoTimGFT, GFTsimplicial}.

As desired, the generalised GFT model reduces to the usual ones
(i.e. to the Barrett-Crane model, in this case) of one drops the
$\theta$s from the vertex function in the action, or equivalently
averages over different orientation data in the same term, {\it
  and} goes to the {\it static-ultra-local} limit of the kinetic term,
i.e. replaces the derivative operators with simple delta functions
identifying pairwise the arguments of the two fields. What we have,
then, is truly a generalisation of the usual formalism, that,
moreover, gives GFTs a structure even more similar to ordinary QFTs,
thanks to the presence of derivatives in the action, and may thus
possibly simplify their future analysis and development.

Other types of GFTs have been constructed in the literature, ranging
from a Boulatov-like model for 3d gravity based on the quantum group
$DSU(2)$ \cite{kirill}, with links to models of 3d quantum gravity
coupled to matter, that we are going to discuss below, to a modified
version \cite{ioetera} of the GFTs for the Barrett-Crane models, with a tunable extra coupling among the
4-simplices and a possible use in the renormalisation of spin foam
models. For all these, and others \cite{grosse}, we refer to the
literature.

\subsection{Matter coupling}
We want instead to discuss now, briefly, the coupling of matter fields to quantum gravity in the group field theory context.
This has been sketched in the 4-dimensional case in
\cite{mikovicmatter}, but has been described in full details only in
the 3D case in \cite{iojimmy}, with an earlier model, based on similar
ideas but covering only the spinless case, having been introduced in
\cite{iojimmylaurent}. The idea followed in this work is to define a
GFT that would treat \lq quanta\rq of gravity and of matter on equal
footing and generate, in perturbative expansion, a sum over Feynman diagrams of gravity, i.e. spin foam
2-complexes/triangulations, and Feynman diagrams of any given matter
field theory, with the latter suitably topologically embedded in the
former. The degrees of freedom of matter fields should also be suitably
coupled to those of geometry to reproduce, at the level of the spin
foam amplitudes, the correct dynamical interaction of gravity and
matter. The spin foam amplitudes describing  this coupling in 3d, were introduced and studied in \cite{PR1}, and
provided the key insight needed to devise an appropriate GFT
reproducing them. The kinematical structure of the new GFT is given
\cite{iojimmy} by {\it two} different types of fields: one is the
usual $\phi$ of the Boulatov theory discussed above, and represents a
triangle with no particle present, the other is a new field
$\psi_s(g_1,g_2,g_3;u): SU(2)^{\otimes 4}\rightarrow \mathbb{C}$ , on
which we impose a global $SU(2)$ symmetry under the simultaneous right
shift of its 4 arguments, and represents a tetrahedron with a single
particle of spin $s$ located at one of its vertices, whose degrees of
freedom are encoded in the group variable $u$. The field $\psi_s$ can
be partially expanded in modes as:
$
\psi_s(g_1,g_2,g_3;u)=\sum_{I,n}\psi^{I}_{sn}(g_1,g_2,g_3)D^{I}_{sn}(u),
$ thus clarifying the interpretation of $s$ as the spin of the
particle and of $I$ as its total angular momentum \cite{PR1}. In the language of
spin networks, the first type of field corresponds to a simple pure
gravity 3-valent spin network vertex, that will give closed spin
network states when suitably contracted with other vertices through
its open links; the second gives a 4-valent vertex with one extra open
link labelled by the particle degrees of freedom; the contraction of
such vertices with both pure gravity vertices and other 4-valent ones
produces {\it open} spin networks, that are at the same
time quantum gravity states and multi-particle states. The GFT should then
describe dynamically the creation/annihilation of these two types of
fundamental structures, producing both a (simplicial)
spacetime and a matter Feynman graph for a particle species with spin
$s$ embedded in it. The mass appears dynamically in the interaction
with gravity, i.e. with the geometric degrees of freedom. For a single
matter field of spin $s$ and in the simplest case of 3-valent
interaction, the full GFT action doing the job is (in a shortened but
intuitive notation):

\begin{eqnarray*} \lefteqn{S[\phi,\psi_s]=\frac{1}{2}\int
\prod^{3}_{i=1}dg_i\;\phi(g_i)\phi(g_i)+\frac{1}{2}\int\prod^{3}_{i=1} dg_i\, du\;
\psi_{s}(g_i;u)\psi_{s}(g_i;u)}\\
&+&\frac{\lambda}{4!}\int\prod^{6}_{i=1}dg_i\;\phi(g_1,g_2,g_3)\phi(g_3,g_5,g_4)\phi(g_4,g_2,g_6)\phi(g_6,g_5,g_1)\\
&+&\mu_2\int \prod^{6}_{i=1}dg_i\, du_a\;
\psi_{s}(g_i,g_3u_a^{-1}hu_a;u_a)\psi_{s}(g_4u_a^{-1}h^{-1}u_a,g_i;u_a)\phi(g_i)\phi(g_i)\\
&+&\mu_3\int \prod^{6}_{i=1}dg_i\, du_a\, du_b\, du_c\;
\psi_{s}(g_i,g_3u_a^{-1}hu_a;u_a)\psi_{s}(g_4u_b^{-1}h^{-1}u_b,g_i;u_b)\psi_{s}(g_i,g_2u_c^{-1}hu_c;u_c)\\
&\times&
\phi(g_6,g_5,g_1)\delta(u_a^{-1}hu_au_b^{-1}h^{-1}u_bu_c^{-1}hu_c)\sum_{I_\alpha, n_\alpha}
D^{I_a}_{sn_a}(u_a)D^{I_b}_{sn_b}(u_b)D^{I_c}_{sn_c}(u_c)C^{I_a\,I_b\,I_c}_{n_an_bn_c}
\end{eqnarray*}
where the third and fourth terms produce, in Feynman expansion,
spacetime building blocks given by tetrahedra with a single
particle propagating along two of its edges (third term) and by
tetrahedra with three particles travelling along three of its
edges and interacting at one of its vertices (fourth term). Here
even the pure propagation of matter degrees of freedom has been
encoded in one of the GFT interaction terms because it involves an
interaction with the geometric degrees of freedom, but it could be
as well encoded in one of the GFT kinetic terms. For a more
extended discussion of this action and for a clarifying graphical
representation of the various terms in it, and of the resulting
Feynman graphs for gravity and matter, we refer to \cite{iojimmy}.
As anticipated, the Feynman amplitudes of this GFT are given by
the coupled Ponzano-Regge spin foam model for gravity coupled to
spinning particles studied in \cite{PR1}, so we do not need to
reproduce them here. These can be understood as discrete path
integrals for quantum gravity on a given spacetime triangulation,
coupled to particles whose trajectories are characterised by the
particle Feynman graph embedded in the same triangulation.
Equivalently, they correspond to the evaluation in the pure
gravity Ponzano-Regge spin foam model of appropriate \lq particle
observables\rq   characterised by the same Feynman graph
\cite{PR1,barrettfeynman}. On the other hand, we want to stress
once more the GFT perspective on the issue of matter coupling in
quantum gravity. This is, again, a purely field-theoretic
perspective, but that of a field theory {\bf of} spacetime with
matter, and not on spacetime. Accordingly, as said, the field
represents at once the basic building blocks {\bf of space and
matter} and the theory, as presently understood, builds up as
Feynman graphs in a perturbative expansion both the histories of
space (possible spacetimes), i.e. of both its geometry and
topology, and of matter (possible particle
evolutions/interactions). Another way in which matter degrees of
freedom may be encoded in a GFT description can be found in
\cite{eterawinston}

The extension of these results to 4D, in both the spin foam and
GFT setting, can proceed in various directions. On the one hand,
one can re-express the Feynman graph amplitudes for any matter
field theory in flat space as a spin foam model for particles
coupled to a topological theory \cite{aristidelaurent4d}, and then
try to build up a full quantum gravity spin foam model reproducing
this in the appropriate topological limit. Having the explicit
form of a such a candidate spin foam model, one could then use it
as a guide for the construction of the corresponding GFT extending
to 4d the model we just presented. On the other hand, the
straightforward extension of the above GFT model to 4D gives a
fully non-perturbative quantum description, in a discrete setting,
of topological gravity (BF theory) coupled to topological strings,
producing in perturbation expansion a sum over both simplicial
spacetimes and discrete string worldsheets embedded in them
\cite{iojimmystrings}, and with kinematical and dynamical
structures consistent with the canonical analysis of such system
performed in \cite{alexjohn}. Yet another direction towards the
understanding of quantum gravity coupled to matter in 4d is the
development of GFT model for gravity coupled to Yang-Mills
theories, following earlier results obtained in the spin foam
setting \cite{iohendryk}, that would produce, again in its
perturbative expansion, a sum over all simplicial complexes with
amplitudes given by the spin foam analogue of a path integral for
gravity coupled to gauge theory, in the form of a lattice gauge
theory on a random simplicial lattice with dynamical geometry
\cite{iojimmyrobert}. We conclude this discussion of
matter-gravity coupling in the GFT language by mentioning a rather
speculative but highly intriguing idea put forward in
\cite{cranematter}. The suggestion is that the GFTs for pure
gravity of the type outlined above already contain all the right
matter degrees of freedom, and therefore there is no need to
modify them by explicitly coupling different types of data and
extending the field content of the models: the idea is that the
conical singularities arising in the GFT perturbative expansion,
and characterising the triangulations not satisfying manifold-like
conditions at their vertices \cite{DP-P}, could be re-expressed as
matter degrees of freedom, moreover possessing apparently the
right type of (internal) structure for reproducing the matter
content of the standard model \cite{cranematter}.

\section{Connections with other approaches}
Where do group field theories stand with respect to other approaches
to quantum gravity? We have already mentioned, albeit briefly, several
links with other types of formalisms, and we would like to
recapitulate them here. In doing so we sketch a (rather speculative at
present) broader picture that sees GFTs
as a {\it generalised formalism for quantum gravity}, in
which other discrete and continuum approaches can be subsumed. We
have already shown the way spin foam models arise universally as
Feynman amplitudes for GFTs, so we don't need to stress again how on
the one hand the GFT approach can reproduce (at least in principle)
any result obtained in the spin foam context, in its perturbative
expansion around the GFT vacuum, while on the other hand providing
much more than that in its, still unexplored, non-perturbative
properties.

In the perspective on GFTs we proposed, they realise a
{\bf local simplicial 3rd quantization of gravity}, with its Feynman
amplitudes interpreted as discrete path integrals for gravity on a
given simplicial complex, and a sum over simplicial spacetimes of all
topologies realised as a perturbative sum over Feynman graphs. The
natural question is then what is the exact relationship with the more
traditional path integral quantisations of simplicial gravity: quantum
Regge calculus (see chapter by Williams) and dynamical triangulations
(see chapter by Ambjorn et al.). Let us consider the first of the
above. This uses a fixed triangulation of spacetime for defining its
quantum amplitudes, and thus should be reproduced at the level of the
GFT Feynman amplitudes for given Feynman graph. The geometric
interpretation of the GFT variables (group representations, group
elements, etc) appearing in the various models, is well understood
\cite{causal, review, alexreview} and was discussed above. In general
one would expect a generic spin foam model for gravity to give a path
integral quantization of discrete gravity in 1st order formalism,
i.e. treating on equal footing the (D-2)-volumes and the corresponding
dihedral angles (equivalently, appropriate parallel transports of a
Lorentz connection) as fundamental variables, as opposed to the 2nd
order formulation of traditional Regge calculus in terms of edge
lengths; but this may be consider a somewhat minor difference, even if
its consequences deserve to be better understood. The main issue that
needs to be solved in order to establish a clear link with the quantum
Regge calculus approach has to do with the fact that the quantum
amplitudes weighting histories of the gravitational field are given,
in the latter approach, by the exponential of the Regge action for
discrete gravity, and in the most studied spin foam models the
connection between the quantum amplitudes and the Regge action is
clear only in a particular regime and even there it is rather
involved, as we have discussed. However, it seems plausible that the
new generalised models of \cite{generalised}, or a suitably
modification of the same, can indeed give amplitudes with the same
structure as in quantum Regge calculus, with a measure being uniquely
determined by the choice of GFT action, thus making the connection
with discrete gravity clear and at the same time subsuming the quantum
Regge calculus approach within the GFT formalism, where it would arise
for a specific type of models, in the perturbative sector only. The
same type of amplitudes is needed also to establish a solid link with
the dynamical triangulations approach, that in fact use again the
exponential of the Regge action for weighting this time not the set of
geometric data assigned to a single
triangulation, but the combinatorial structure of the triangulation
itself, which is treated as the only true dynamical variable within a
sum over all possible triangulations of a given topology. Once the
right type of amplitudes is obtained, the dynamical traingulations
approach would once more arise as a subsector of the GFT formalism, if
one could find the right
way of trivialising the extra structure associated to each
triangulation, thus dropping the now redundant sum over geometric
data. Of course, more work would be needed then to impose the
extra conditions (fixed slicing structure, absence of baby universe
nucleation, etc) that seem to be needed in the modern version of the
approach (see chapter by Ambjorn et al.) to have the right type of
continuum limit. Work on this is in progress \cite{GFTsimplicial}.

\

That a covariant path integral quantisation can be understood as being
more general than the corresponding canonical/Hamiltonian one, and
that this is even more true when a sum over topologies is implemented
through some sort of 3rd quantisation, is well known. In the GFT
framework, one expects then to be able to reproduce the setting and
results of a canonical quantum gravity based on the same type of
variables (group elements, interpreted as holonomies of a
gravitational connection, or group representations, interpreted as
(D-2)-volumes conjugate to them), and of quantum gravity states (spin
networks): loop quantum gravity. How this can indeed be realised
\cite{laurentgft}, in a way that goes even beyond the present results
of the LQG approach, has been shown above, when discussing the GFT
definition of the canonical inner product. The main differences
between the particular version of the LQG formalism that the GFT
approach reproduces, and the traditional one (see chapter by
Thiemann), are two: 1) the spin networks appearing as boundary states
or observables in the GFT framework are inherently adapted to a
simplicial context in that they are always D-valent in D spacetime
dimensions, being dual to appropriate (D-1)-triangulations, while the
spin networks arising in the {\it continuum} loop quantum gravity
approach are of arbitrary valence; 2) the group used to label these
states and their quantum evolution amplitudes in the GFT case is the
Lorentz group of the corresponding dimension, which means, in
dimension 4 and minkowskian signature, the non-compact group
$SO(3,1)$, while LQG uses $SU(2)$ spin networks also in this
context. The first of these differences is not so crucial, since on
the one hand any higher valent spin network in LQG can be decomposed
into lower-valent ones, and on the other hand any coarse graining
procedure to be implemented to approximate
simplicial (boundary) structures with continuum ones, and applied to
the GFT boundary states or to GFT observables, would likely remove any
restriction on the valence. The second difference is more troubleful,
and establishing an explicit connection between the fully covariant
GFT spin network structures and the $SU(2)$-based LQG ones is no easy
task. However, lots of work has already been done on this issue
\cite{eterasergei} (see
chapter by Livine) and can be the starting point for 1) establishing a well-defined canonical
formalism from the GFT structures first, and then 2) linking (more
appropriately, reducing, probably through some sort of gauge fixing)
this formalism to that of traditional LQG.

\

A fourth approach that can be linked to the GFT one is the causal set
approach (see chapter by Henson). Recent work on spin foam models and
GFT \cite{causal, feynman, generalised} has shown how the GFT Feynman
amplitudes can be written in the form of models for causal evolution
of spin networks \cite{fotinilee}, by a correct implementation of
causality requirements. A key step in doing this is the causal
interpretation, in the Lorentzian context, of the GFT Feynman graph,
this being {\it directed graph} i.e. a graph with \lq directions\rq or
arrows labelling its edges thus endowed with an orientation. In this
interpretation, the vertices of the graph, i.e. the elementary GFT
interactions, dual to D-simplices, are the fundamental spacetime
events, and the links of the graph each connecting two such vertices,
dual to (D-1)-simplices and corresponding to elementary propagation of
degrees of freedom in the GFTs, represent the fundamental causal
relations between spacetime events. A directed graph is very close to
a causal set, from which it differs for just one, albeit important,
property: it possibly includes closed loops of arrows. This, from the point of view of causal set theory, is a
violation of causality, the microscopic discrete equivalent of a
closed timelike loop in General Relativity, forbidden in the basic
axioms defining the approach. No such restriction is imposed, a
priori, on the corresponding GFT structures. There are several
possible attitudes towards this issue from the GFT perspective: 1) it
is possible that such configurations, even though they are present in
the set of allowed configurations, are not relevant for the continuum
approximation of them, i.e. they disappear or give a negligible
contribution to the sum under the appropriate coarse graining
procedure; 2) in the specific GFT models that will turn out to be of
most interest for quantum gravity, Feynman graphs possessing such \lq
closed timelike loops\rq   may end up being assigned quantum amplitudes
that strongly suppress them compared to other configurations; 3) one
may be able to give a purely field-theoretic interpretation of such
loops in the GFT context and then identify some sort of \lq
superselection rules\rq   that could eventually be imposed on the GFT
perturbative expansion to suppress them; 4) finally, one may decide
that there is no fundamental reason to ban such configuration from the
admissible ones and rather find the way to interpret them physically
and study their observable consequences. Only further work will
tell. Finally, one more difference with the causal set framework is
worth mentioning: due again to the simplicial setting in which they
are realised, the GFT Feynman graphs have vertices of finite and fixed
valence depending on the spacetime dimension, while the causal set
vertices have no restriction on their valence. Once more, it is well
possible that one has to welcome such restriction because it results
in may be interpreted as one more sign of a fundamental spacetime
discreteness, that may be attractive from both
philosophical and physical reasons; at the same time, it is possible
that such restriction on valence will be removed automatically and
necessarily in the study of the continuum approximation of such
discrete substratum for spacetime, by means of coarse graining
procedures (as mentioned above regarding GFT kinematical states) or of
renormalisation group arguments (e.g. inclusion of more types of
interaction terms in the GFT action).

\

The GFT formalism is therefore able to encompass several other
approaches to quantum gravity, each carrying its own set of ideas and
techniques; strengthening the links with these other approaches will
be, in our opinion, of great importance for both developing further
the GFT framework itself, and also for making progress on the various
open issues that such other approaches still have to face, thanks to
their \lq embedding\rq   into a different context, that naturally brings
in a fresh new perspective on the same open issues.

\section{Outlook}
Let us summarise. The group field theory approach aims to describe the
dynamics of both spacetime geometry and topology down to the Planck
scale, in a background independent and non-perturbative way (even if
at present almost only the perturbation expansion around the \lq
complete vacuum\rq   is well understood), using a field-theoretic
formalism. In essence, as discussed, a group field theory is a field
theory over a group manifold, as for the mathematical formulation, and
at the same time a field theory over a simplicial superspace (space of
geometries), as for the physical interpretation. Thanks to the
discrete description of geometry that the simplicial setting allows
for, it corresponds to a {\it local} 3rd quantisation of gravity, in
which the \lq quanta\rq   being created and annihilated in the quantum
evolution are not universes, as in the traditional approach, but
appropriately defined chunks of space.

\

What is particularly attractive, in our opinion, about this approach
is the combination of {\it orthodoxy} in the mathematical language and
technology used and of {\it radicalness} in the ideas that this
language expresses. On the one hand, in fact, GFTs are almost ordinary
field theories, defined on a group manifold with fixed metric and
topology, and thus background dependent, speaking from a purely formal
point of view, and using the ordinary language of fields and actions,
of Feynman graphs, propagators and vertices of interaction, gauge
symmetries, etc. This means that GFTs allow, at least in principle, to
tackle any of the traditional questions in quantum gravity using
techniques and ideas from QFT, thus making use of the vast body of
knowledge and methods developed in a background dependent context that
appeared for long time not directly applicable to quantum gravity
research.
On the other hand, the overall picture of spacetime and of gravity
that this approach is based on, despite the traditional nature of the
language used to express it, is definitely radical and suggests the
following. There exist fundamental building blocks or atoms of space,
which can be combined in all sorts of ways and can give rise to all
sorts of geometry and topology of space. At the perturbative level
spacetime is the (virtual) history of creation/annihilation of these
fundamental atoms; it has no {\it real} existence, at least no more
real existence in itself than each of the infinite possible
interaction processes corresponding to individual Feynman graphs in
any field theory; the interaction/evolution of these building blocks
does not leave neither the geometry nor the topology of space fixed,
but treats them on almost equal footing as dynamically evolving; the
description of this evolution is necessarily background independent
(from the point of view of spacetime) because spacetime itself is
built from the bottom up and all of spacetime information has to be
reconstructed from the information carried by the \lq atoms\rq and
thus by the Feynman graphs. At the non-perturbative level, for what we
can see given the present status of the subject, spacetime is simply
not there, given that the non-perturbative properties of quantum
gravity would be encoded necessarily either in the GFT action, and in
the resulting equations of motion, or in the GFT partition function,
and the related correlation functions, to be studied
non-perturbatively, neither of which need any notion of spacetime to
be defined or analysed (through instantonic calculations, or
renormalisation group methods, or the like). Spacetime information is
thus necessarily encoded in structures that do not use {\it per se} a
notion of spacetime. Finally, there would be a {\it fundamental
  discreteness} of spacetime and a key role played by {\it causality},
in the pre-geometric sense of {\it ordering} related to the notion of
orientation (so that it would probably be better to talk about \lq pre-causality\rq).

Many of these ideas had been of course put forward several times in
the past, and occur in more than one other approach to quantum gravity,
but the GFT formalism brings all of them together within a unique
framework and, as said, expresses them in a rather conventional and
powerful language; moreover, thanks to the possibility (yet to be
realised) of subsuming many other approaches within the GFT one, all
the ideas and techniques developed and the results obtained within these other
approaches will maybe find a new role and application in the GFT
setting.

\

Let us sketch also some examples of how traditional field
theoretic methods can be used to tackle within a radically new
perspective some crucial open issue in quantum gravity research.
We have already mentioned several of these examples: the
long-standing problem of solving the Hamiltonian constraint
equation of canonical quantum gravity can be identified with the
task of solving the classical GFT equations of motion, and how the
other long-standing issue of defining a canonical inner product
for quantum gravity states is turned into the task of analysing
the tree level truncation of the (perturbative expansion of the)
appropriate GFT. Also, the perturbation theory around such quantum
gravity states would be governed, according to the above results,
by the approximation of the GFT partition function around its
classical solutions, and this suggests a new strategy for
investigating the existence of gravitons (propagating degrees of
freedom) in specific GFT/spin foam models. The most outstanding
open issue that most of the discrete non-perturbative approaches
to quantum gravity still face is however that of the continuum
approximation. This problem has been formulated and tackled in a
variety of ways, depending on the particular approach to quantum
gravity under scrutiny. Obviously, given the role that formalisms
like dynamical triangulations, quantum Regge calculus, causal sets
or loop quantum gravity can play within the group field theory
framework, all the various techniques developed for them can be
adapted to the GFTs. However, the field theory language that is at
the forefront of the GFT approach suggests once more new
perspectives. Let us sketch them briefly. The continuum
approximation issue can be seen as the search for the answer to
two different types of question. a) What is the best procedure to
approximate a discrete spacetime, e.g. a simplicial complex, with
a continuum manifold, and to obtain some effective quantum
amplitude for each geometric configuration from the underlying
fundamental discrete model? In the context of spin foam models,
the most directly linked with the GFT approach, this amounts to
devising a background independent procedure for \lq coarse
graining\rq the spin foam 2-complexes and the corresponding
amplitudes \cite{robert,fotini} to obtain a smooth approximation
of the same. b) If a continuum spacetime or space are nothing else
than some sort of \lq condensate\rq of fundamentally discrete
objects, as in some \lq emergent gravity\rq approaches (see
chapter by Dreyer), and as suggested by condensed matter analog
models of gravity \cite{volovik, analog}, what are these
fundamental constituents? what are their properties? what kind of
(necessarily background independent) model can describe them and
the whole process of \lq condensation\rq? What are the effective
hydrodynamic variables and what is their dynamics in this \lq\lq
condensed or fluid phase\rq\rq ? How does it compare to GR?

For what concerns the first (set of) question(s), the GFT approach
offers a potentially decisive reinterpretation: since spin foam
are nothing else than Feynman graphs of a GFT, and that spin foam
models are nothing else than their corresponding Feynman
amplitudes, the coarse graining of a spin foam model, either
performed as outlined in \cite{robert} or with the techniques
introduced in \cite{fotini}, corresponds exactly to the {\it
perturbative renormalisation} of the corresponding GFT. On the one
hand this suggests to deal with the problem of continuum
approximation of spin foams using all the perturbative and
non-perturbative renormalisation group techniques from ordinary
field theory adapted to the GFT case. On the other hand gives a
further justification for the idea, proposed in \cite{fotini},
that the Connes-Kreimer Hopf algebra of renormalisation developed
for QFT could be the right type of formalism to use in such a
quantum gravity context.

As for the second (set of) question(s), the GFT approach
identifies uniquely the basic building blocks of a quantum space,
those that could be responsible for the kind of \lq
condensation\rq process or the transition to a fluid phase at the
root of the emergence of a smooth spacetime in some approximation
and physical regime, and gives a precise prescription for their
classical and quantum dynamics, that can now be investigated. In
particular, one could develop a statistical mechanics picture for
the dynamics of the GFT \lq atoms\rq of space, and then the above
idea of a \lq condensation\rq  or in general of the possibility of
an hydrodynamic description could be tested in specific GFT
models, and in very concrete and precise terms, using once more
usual techniques from statistical field theory (as applied for
example in condensed matter systems). A more detailed discussion
of the possible development of GFTs along these lines can be found
in \cite{ioGFTemergence}.

\
Whether any of the above ideas will be realised, and whether other,
not yet imagined, possibilities for new developments will become
manifest in the near future, only further work will tell. In our
opinion, however, it is already clear that the GFT approach can be the
right framework for asking the most fundamental questions about
Quantum Gravity, and for finding answers to them.

\end{document}